\documentclass[twocolumn,showpacs,preprintnumbers,amsmath,amssymb,prl]{revtex4-1}

\usepackage{color}\definecolor{gray}{rgb}{0.5,0.5,0.5}

\usepackage{graphicx,textcomp}
\usepackage{dcolumn}
\usepackage{bm}
\usepackage{color}
\usepackage[utf8]{inputenc}
\usepackage[T1]{fontenc}
\usepackage{soul}

\begin{document}

\title{Dopant induced ignition of helium nanodroplets in intense few-cycle laser pulses}

\author{S. R. Krishnan$^1$}
\author{L. Fechner$^2$}
\author{M. Kremer$^1$}
\author{V. Sharma$^1$}
\author{B. Fischer$^1$}
\author{N. Camus$^1$}
\author{J. Jha$^{4}$}
\author{M. Krishnamurthy$^{4}$}
\author{T. Pfeifer$^{1}$}
\author{R. Moshammer$^{1,3}$}
\author{J. Ullrich$^{1,3}$}
\author{F. Stienkemeier$^2$}
\author{M. Mudrich$^2$}
\affiliation{$^1$Max-Planck-Institut f{\"u}r Kernphysik, 69117 Heidelberg, Germany}
\affiliation{$^2$Physikalisches Institut, Universit\"at Freiburg, 79104 Freiburg, Germany}
\affiliation{$^3$Max Planck Advanced Study Group at CFEL, Luruper Chaussee 149, 22761 Hamburg, Germany}
\affiliation{$^4$Tata Institute of Fundamental Research, 1 Homi Bhabha road, Mumbai 400005, India}
\author{A. Mikaberidze$^3$}
\author{U. Saalmann$^{3,5}$}
\author{J. -M. Rost$^{3,5}$}
\affiliation{$^5$Max-Planck-Institut f{\"u}r Physik komplexer Systeme,
             N\"othnitzer Str. 38, 01187 Dresden, Germany}


\date{\today}

\begin{abstract}
\noindent
We demonstrate ultrafast resonant energy absorption of  rare-gas doped
He nanodroplets from intense 
few-cycle ($\sim$10\,fs)  laser pulses. We find that
less than 10 dopant atoms ``ignite'' the  droplet to generate a non-spherical electronic nanoplasma resulting ultimately in  complete ionization and disintegration of all atoms, although the pristine He droplet is transparent for the laser intensities applied. Our calculations at those intensities reveal that the minimal pulse length required for ignition is about 9\,fs.

\end{abstract}

\maketitle

\noindent
The ionization dynamics of atomic clusters in intense ultrashort laser pulses has been an active area of research in recent years at near-infrared (NIR), vacuum-ultraviolet (VUV) and soft X-ray wavelengths \cite{sasi06,feme10,wabi02}. The energy absorption of NIR light by rare-gas clusters is far more efficient
as compared to atomic jets or planar solid targets when irradiated with similar laser pulses.
The underlying mechanism is resonance absorption by an electronic nanoplasma which forms through illumination with the light \cite{dido96,saro03,sa06}. The resonance occurs when the background ions of total charge $Q$  occupy a large enough sphere of radius $R$ such that  the nanoplasma  eigenfrequency $\Omega = \sqrt{Q/R^{3}}$
 matches the frequency $\omega$ of the laser light
 \cite{sasi06}. The resonance condition $\Omega = \omega$ is achieved on the time scale of atomic motion (sub- or few-picoseconds) since the cluster, i.e. the ions, must {\it expand} in order for the ions to  become sufficiently dilute. This  has been verified in experiments where pump-probe delay \cite{zwdi99} or pulse durations are varied \cite{dofe05}, as well as in calculations \cite{saro03}.

The fact that this resonance absorption at NIR frequencies relies predominantly on atomic expansion has two consequences: Firstly, it requires time scales of ionic motion in spite of its electronic nature. Secondly, it does not occur in isolated atoms or those inside the bulk in condensed phase.
One may ask if there is another possibility for this  resonant coupling of energy from the light pulse to the electrons which is not only extremely efficient but also fast,   bypassing atomic expansion. Indeed this can be achieved as theoretically proposed and demonstrated \cite{misa09}, if one uses a two component system with a few seed atoms with ionization potentials lower than the majority of atoms which belong to the other component.
If the spherical symmetry of the system is suitably broken (e.g. Xe atoms in the center of a He droplet irradiated by linearly polarized light)  a ``cigar shaped'' nanoplasma is formed. It has two eigenfrequencies which bracket the eigenfrequency of a spherical plasma. The lower one (along the linear laser polarization) is  resonant with 790 nm laser light right away at typical atomic cluster densities leading to almost immediate resonance absorption without the need for atomic expansion. This allows not only to realize the efficient energy transfer very quickly, now limited essentially by the rise time of the laser pulse, but also opens the resonant light absorption to other  forms of matter which do not (Coulomb) explode.

Here, we report on experiments which demonstrate this purely electronic resonance absorption with rare-gas doped helium droplets illuminated by a few-cycle pulse of a duration as short as 10\,fs at 790\,nm. This rules out the influence of any kind of atomic motion. Reducing the pulse length even further must eventually prevent even this electronic resonance absorption. Indeed, as we will demonstrate with calculations,  there is a clear transition from dopant induced resonance absorption to a regime dominated by static field ionization \cite{gnsa09} if the pulse length falls below a critical value which depends of course on the intensity of the laser pulse.

%
%

He nanodroplets can be doped with a well-controlled number of rare-gas atoms which aggregate at the center of the droplet \cite{tovi04,stle06}. These are exceptionally suitable to demonstrate the role of seed atoms in resonance absorption, since the laser intensity can be chosen such that the light couples almost exclusively to the dopants whereas the pristine He droplet is transparent. Interestingly this applies to NIR \cite{misa08,misa09} as well as to X-ray \cite{gnsa09} frequencies. Despite its transparency, the droplet becomes highly active once the core of seed atoms is ionized.

In the experiment a beam of He nanodroplets is produced by expanding pressurized $^4$He gas (70-90 bar) through a nozzle ($5\,\mu$m in diameter) maintained at a temperature of 15-25\,K. By varying the nozzle temperature in this range the mean number of He atoms per droplet is adjusted in the range 10$^3$-10$^5$. Owing to their low internal temperature $\sim 370\,$mK, these droplets are in a superfluid state \cite{tovi04, stle06}. They are  doped by passing the skimmed beam of pure droplets through a cylindrical doping cell that is 3\,cm long with collinear apertures (\O\,=\,3\,mm). The mean number of dopants per nanodroplet $K$ is regulated through controlled leaking of the desired rare-gas  into this cell using a dosing valve (leak rate $< 10^{-10}$\,mbar\,l/s). The pressure in the doping cell is monitored by a directly attached vacuum gauge.

Recently, we have simulated the pick-up process of alkali atoms that form high-spin clusters using a Monte-Carlo model \cite{bust11}. Here, we use the same model adapted to rare-gas dopants to validate the estimate of doping levels and to estimate the loss of He due to evaporation caused by the doping process: The number $K$ of dopant atoms
is ascertained from the cell pressure and the droplet size according to the modified Poissonian pick-up statistics \cite{stle06} using the semi-empirical formula of Kuma et al. \cite{kugo07} which takes evaporation of He atoms into account.

\begin{figure}[t!]
\begin{center}
\includegraphics[width=0.48\textwidth]{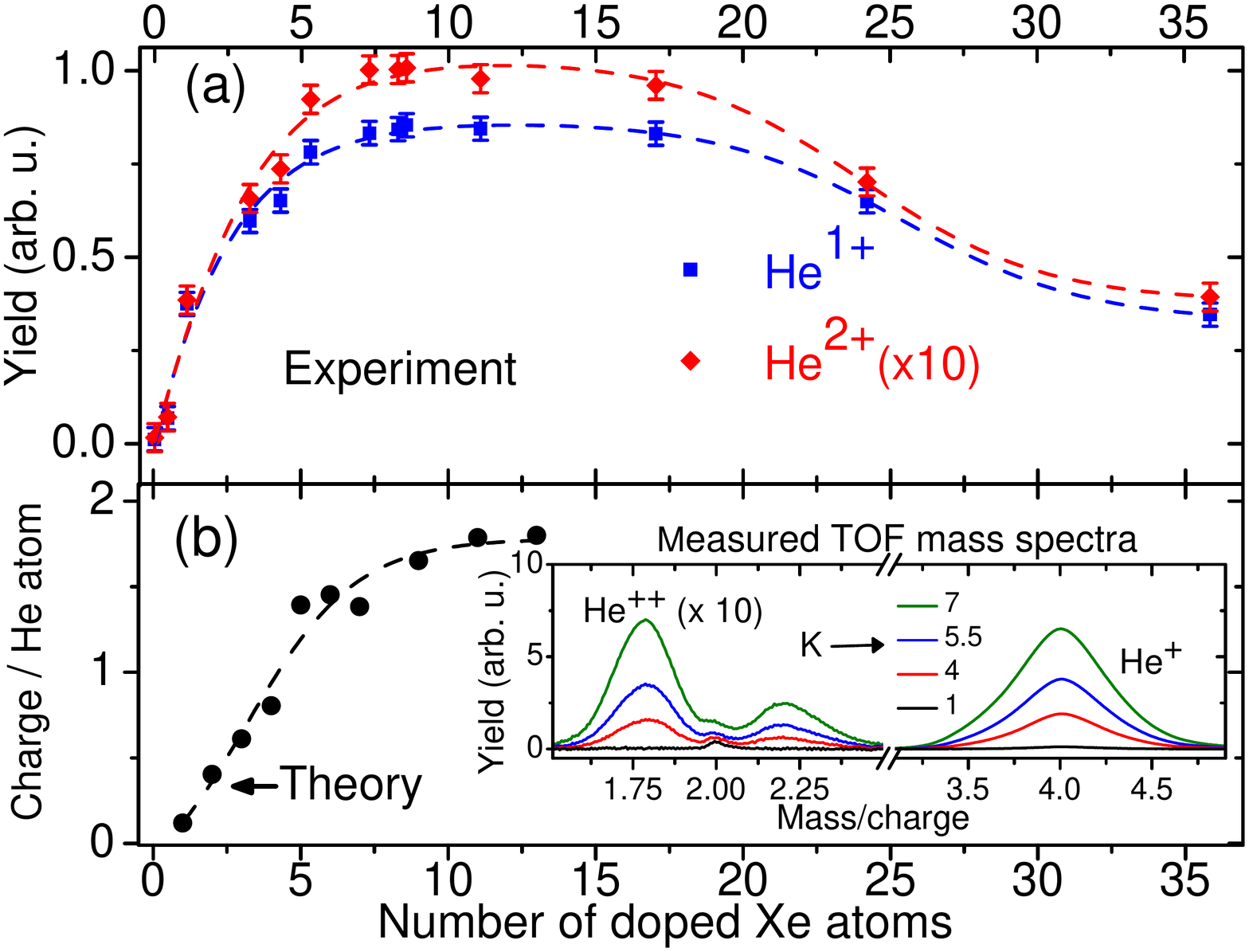}
\caption{(a) Yields of He$^{++}$ and He$^{+}$ ions as a function of the mean number of Xe dopants in a nanodroplet containing $1.5{\times}10^4$ He atoms at a peak laser intensity of $7{\times} 10^{14}$\,W/cm$^{2}$. 
(b)  Numerical calculation of charge per He atom as a function of number of Xe atoms in a droplet containing 4000 He atoms. All lines are to guide the eye. Inset: Time-of-flight mass spectra of He$^{++}$ and He$^{+}$ ions for different numbers of dopant Xe atoms ($K$) as shown in the legend.
\label{fig:HepHe2p} }
\end{center}
\end{figure}

Intense few-cycle laser pulses ($\sim$10\,fs) at a central wavelength of 790\,nm with peak intensities in the range $10^{14}$--$10^{15}$\,W/cm$^{2}$ are generated from a Ti-Sapphire based mode-locked laser system. Photo-ions are detected by a time-of-flight (TOF) spectrometer in the Wiley-McLaren geometry. TOF mass spectra of He$^{++}$ ions for different mean doping numbers $K$ of Xe in a droplet containing $1.5{\times}10^4$ He atoms are shown in the inset in Fig.~\ref{fig:HepHe2p}\,(b). The He$^{++}$ peak, which is characteristic for ionizing doped He droplets and is absent in mass spectra of atomic He gas, is split  as a result of the finite angular acceptance of the spectrometer into two components due to ions with high momenta directed towards and away from the detector, respectively.
The yield of He$^{++}$ and He$^+$ ions is extracted from the TOF mass spectra by integrating over the respective mass peaks.

\begin{figure}[t!]
\begin{center}
\includegraphics[width=0.48\textwidth]{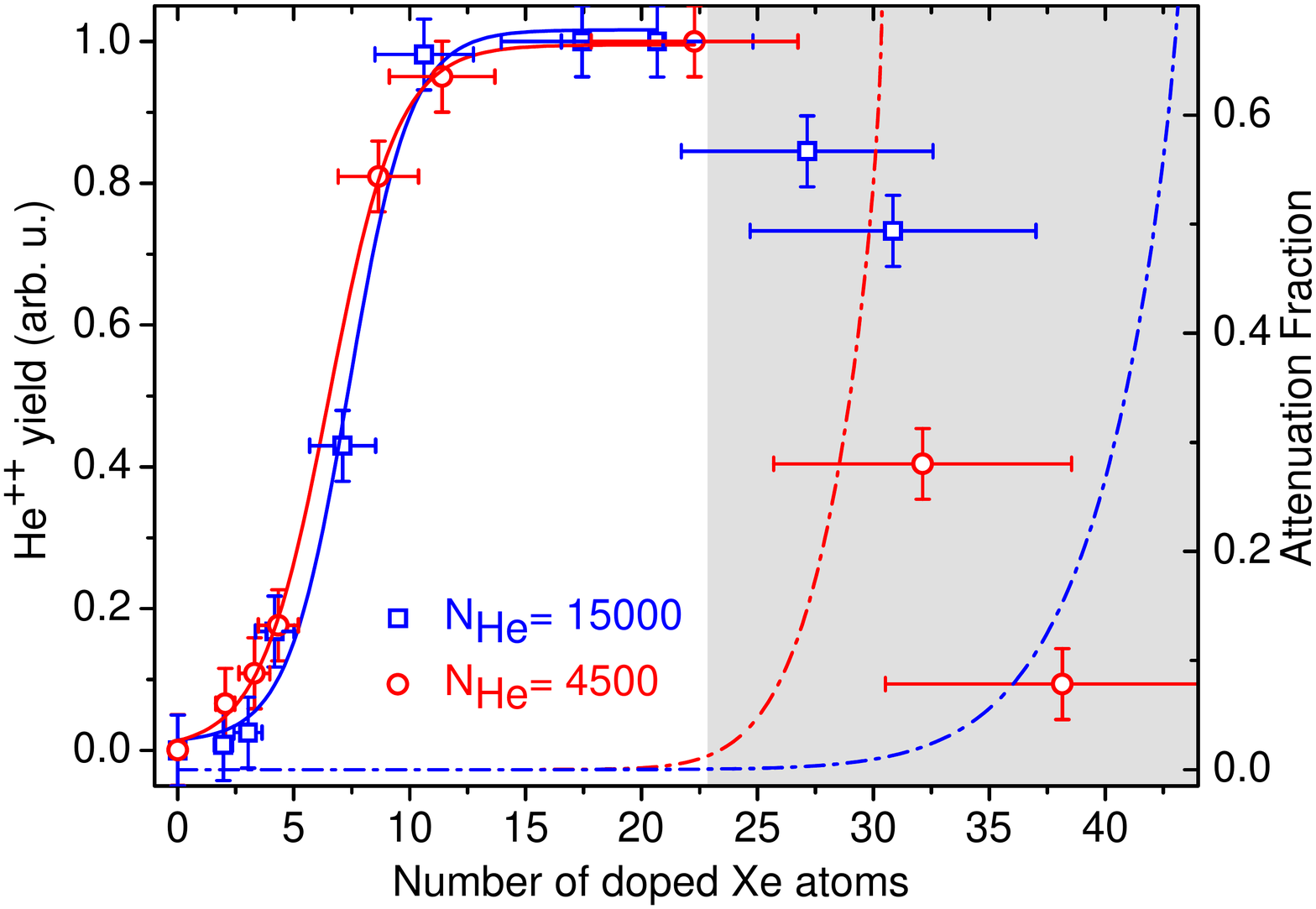}
\caption{Xe doping dependence of He$^{++}$ yields at a peak laser intensity of $1.5\times10^{14}$\,W/cm$^{2}$ for droplets with a mean number of He atoms per droplet $N_{He}$=4500(red) and $N_{He}$=15000(blue). Horizontal bars present the error in estimating the number of doped Xe atoms. The dash-dot curves present the fraction of droplets lost due to complete evaporation, which is significant only for number of Xe dopant atoms $K\gtrapprox 20$ (grey shading).
\label{fig:Sizes} }
\end{center}
\end{figure}

Fig.\ref{fig:HepHe2p}\,(a) presents the doping dependence of the yields of He$^{++}$ and He${^+}$ at a peak intensity of $7\times 10^{14}$\,W/cm$^{2}$ of the laser pulse. A gradual increase of doping number $K=1$ up to 10 leads to a dramatic step-like increase in the yields of He$^{+}$ and He$^{++}$ ions. We refer to this nanoplasma formation as dopant-induced ignition (DII). The saturation of ion yields, and equivalently the build-up of charge in the nanoplasma, occurs for a critical doping number $K_\mathrm{cr}$ of just $\sim$7(\textpm1.4). The ionization of the He droplet is not enhanced any further by adding more dopants.
Based on classical molecular dynamics (MD) \cite{saro03, sa06}  we have tailored a theoretical description to the present problem of doped He droplets (for details of the approach see \cite{misa09,misa08}).
The result from our calculation
shows (Fig.\ref{fig:HepHe2p}\,(b)) qualitatively the same behavior, although a direct comparison of the charge to yields is difficult since the experiment is an average over the droplet size and doping distributions as well as over intensities in the focal volume. Moreover, in the theoretical result electron recombination is not taken into account.
Yet, the curves agree well qualitatively and even quantitatively regarding the critical dopant number at which the saturation of charging occurs. This agreement points to a robust underlying mechanism.

\begin{figure}[t!]
\begin{center}
\includegraphics[width=0.48\textwidth]{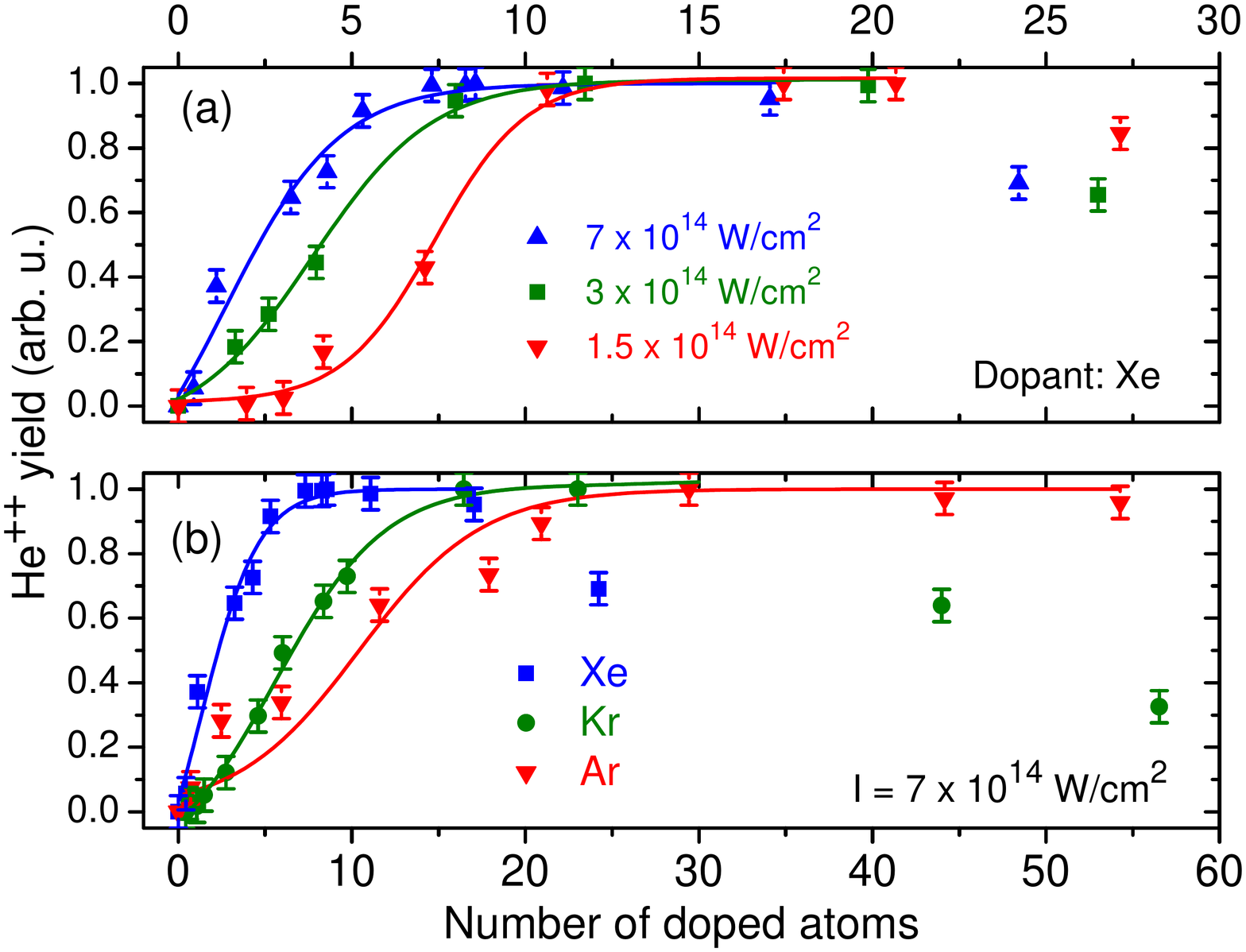}
\caption{The He$^{++}$ yields for a droplet containing 15000 He atoms as function of the number of Xe dopant atoms (a) for different laser intensities  $I$  and (b) for different dopant species at $I = 7\times 10^{14}$\,W/cm$^{2}$.
\label{fig:Intens_Elem} }
\end{center}
\end{figure}

The critical doping number is remarkably independent of the mean droplet size as shown in Fig.~\ref{fig:Sizes} having the value of 10(\textpm2) for mean droplet sizes of 4500 and 15000 atoms at a peak laser intensity of $1.5\times 10^{14}$\,W/cm$^{2}$. The error in estimating the number of dopants using Kuma's approach \cite{kugo07} is also shown.
The decrease of doubly charged He for increasing $K$ in Figs.\,\ref{fig:HepHe2p}--\ref{fig:Intens_Elem}
is related to the doping process: When dopant atoms are picked up sequentially and aggregate inside the droplet, collisional as well as binding energy is released into the droplet. This leads to an increasing fraction of He droplets that completely evaporate at high doping pressure. The corresponding estimate from the Monte-Carlo simulation for Xe doping is indicated in Fig.~\ref{fig:Sizes} as dashed lines. Clearly, the onset of significant nanodroplet destruction in this case of Xe doping occurs well beyond the $K_{cr}$\,($\sim10$) observed in the experiment. Thus, the process of dopant-induced ignition  is not influenced by evaporation effects.

The insensitivity of DII with respect to the droplet sizes  results simply from the fact that
DII sets in with the formation of a cigar shaped plasma which is much smaller than the droplet itself \cite{misa09}.
In contrast, DII should depend on the laser intensity and similarly on the ionization potential of the seed atoms, as well as on the pulse length.
The former dependences are illustrated with experimental results
 in Fig.~\ref{fig:Intens_Elem}. As expected, $K_\mathrm{cr}$ is smaller for higher intensities and/or lower ionization potentials.

\begin{figure}[t!]
\begin{center}
\includegraphics[width=0.45\textwidth]{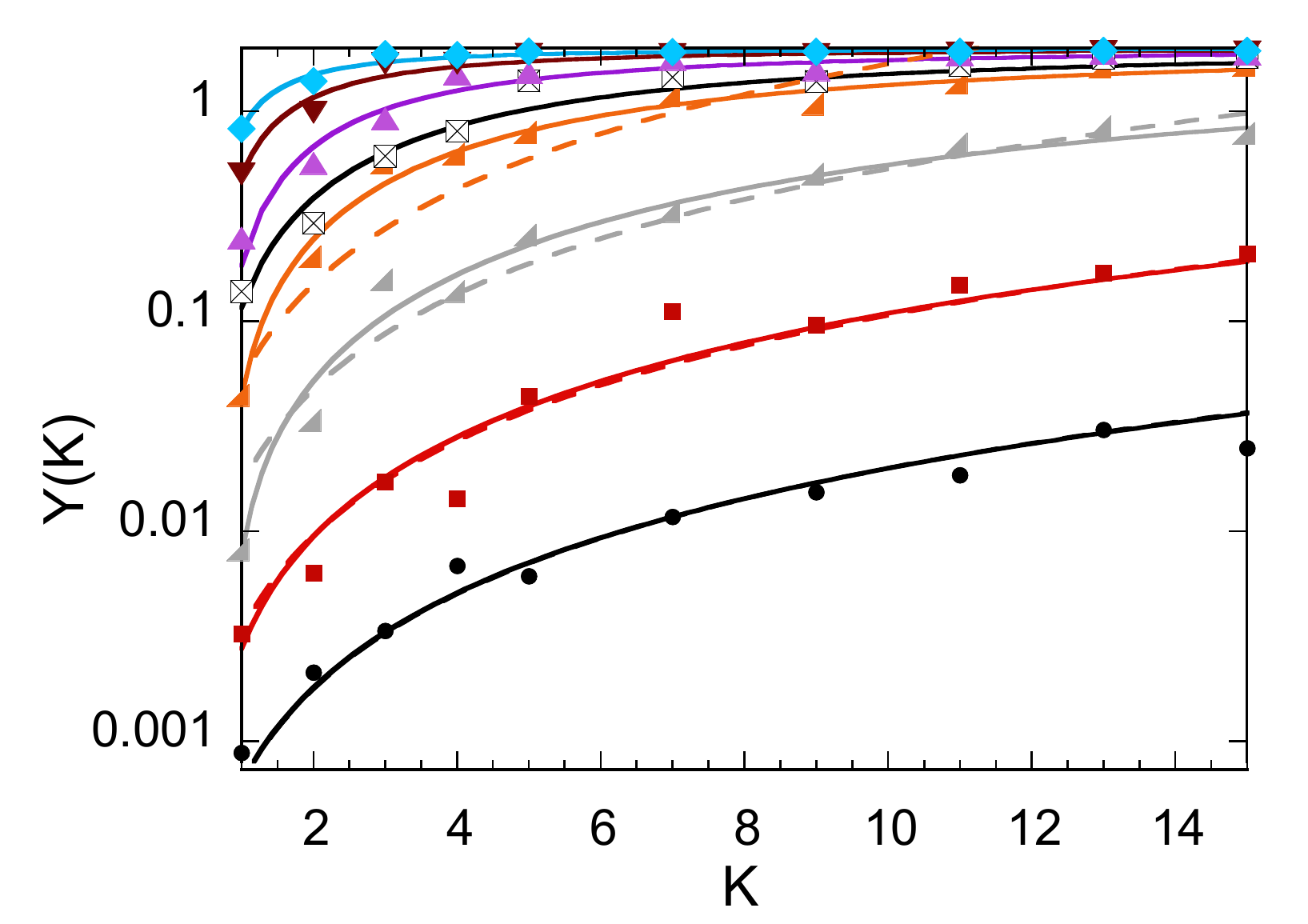}
\caption{Average charge per He atom $Y(K)$ as a function of the number of doped Xe atoms for a droplet with 4000 He atoms and intensity as in Fig.~\ref{fig:HepHe2p} from MD calculations. The different curves distinguish from bottom to top
laser pulse lengths (half width half maximum) from 3 to 15\,fs in steps of 2\,fs. Additionally, $Y(K)$ for 10\,fs (crossed boxes) is shown, see also Fig.~\ref{fig:HepHe2p}. The  lines
are fits of the data points according to Eq.\,(\ref{shape})
(solid) and with
 the power law  $Y \propto K^{3/2}$ corresponding to  ionization by a static field for $T \le 9$\,fs (dashed). For details, see text.
 \label{fig:PulseLength}}
\end{center}
\end{figure}

We determined numerically the average He charge in a droplet $Y(K)$ as a function of the number of Xe dopants $K$. One can see in Fig.~\ref{fig:PulseLength} that at the intensity of $7\times\,10^{14}$\,W/cm$^{2}$  a qualitative change of the average charge per He atom $Y(K)$  occurs between 7 and 9\,fs pulse length: For longer pulses DII (solid lines) is operative with a fast rise  followed by saturation at full ionization of the He atoms ($Y=2$).
   For shorter pulse lengths the average charge  shows a gradual rise $Y \propto K^{3/2}$ (dashed lines) which is characteristic for  ionization due to the static electric field of the laser ionized Xe ions in the center of the droplet.

 To derive this dependence  we
assume that all Xe ions constitute a point charge at
the center of the droplet and that the  density of He atoms is
uniform within the spherical droplet. Then the problem is of purely radial nature and
a sphere  exists which contains only doubly ionized helium atoms. Its radius is given by
 $R_{++} =
\sqrt{q_\mathrm{Xe} K/E_\mathrm{bs2}}$ where $q_\mathrm{Xe}$ is the average charge of each of the $K$ dopant atoms and
$E_\mathrm{bs2}$ is the electric field necessary to ionize the second electron in He through barrier suppression (Bethe rule) \cite{besa57,gnsa09}.
The number of He atoms inside this sphere is given by
$N_{++} = 4 \pi \rho_\mathrm{He} R_{++}^3/3 = 4 \pi
\rho_\mathrm{He} (q_\mathrm{Xe} K/E_\mathrm{bs2})^{3/2} /3$, where
$\rho_\mathrm{He}$ is the density of He in the droplet. We find the number of
singly ionized helium atoms $N_{+}$ in a similar way and obtain
the average ion charge per He atom using $Y = ( N_{+} + 2
N_{++})/N_\mathrm{He}$, where $N_\mathrm{He}$ is the total
number of He atoms in the droplet. This yields
\begin{equation}\label{eq:qhe-fi}
Y(K) = \frac{4 \pi \rho_\mathrm{He}}{3 N_\mathrm{He}} \left(
 E_\mathrm{bs1}^{-3/2} + E_\mathrm{bs2}^{-3/2} \right) q_
\mathrm{Xe}^{3/2} K^{3/2},
\end{equation}
%
where $E_\mathrm{bs1}$, $E_\mathrm{bs2}$ are the first and the second
barrier suppression fields for He. Assuming that $q_\mathrm{Xe}$
does not depend strongly on the number of Xe atoms $K$ (this is
fulfilled very well for $K$ between 2 and 15 as seen from the MD
simulations), we conclude that $Y \propto K^{3/2}$,
i.\,e. the static field ionization due to positively charged dopants
follows a characteristic power law.

\begin{figure}[t!]
\begin{center}
\includegraphics[width=0.4\textwidth]{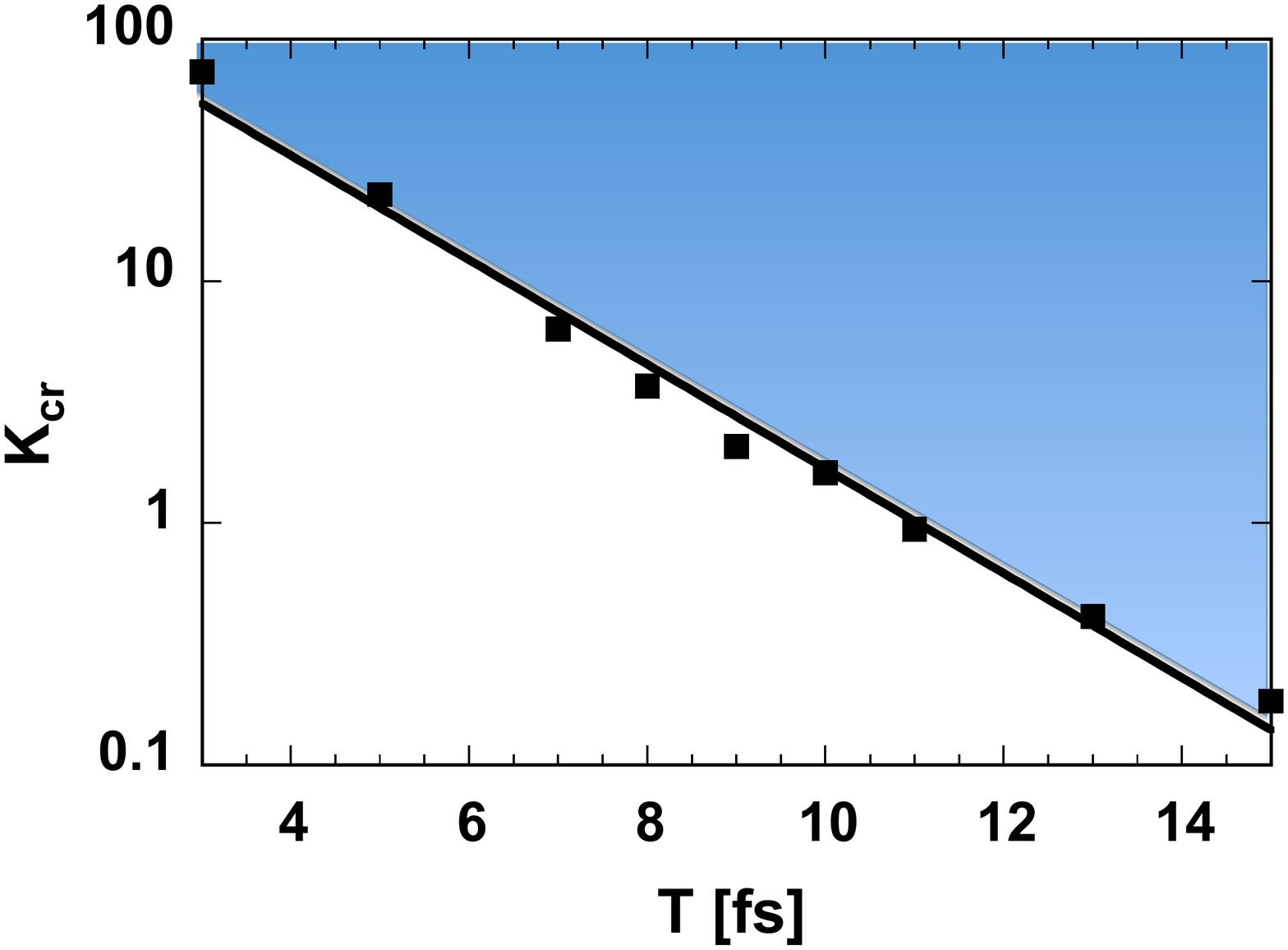}
\caption{The number of dopants for which the average charge per He atom $Y$ in Fig.~\ref{fig:PulseLength} has a turning point, $Y''(K_\mathrm{cr})=0$. The shaded area distinguishes the regime of dopant induced ignition (DII) from to the field ionization regime (not shaded), at a peak intensity of $7\times 10^{14}$\,W/cm$^{2}$.
\label{fig:field2ignition}}
\end{center}
\end{figure}

Although it is clear from Fig.~\ref{fig:PulseLength} that for increasing pulse lengths
the average charge per He atom $Y(K)$ is no longer  well described by the field ionization power law (dashed lines),
a well defined transition from field ionization to DII  for increasing pulse length is difficult to extract.  To determine the   minimal pulse length $T$ for which DII dominates, we introduce a function for the shape of $Y(K)$, which can describe both limits,
\begin{equation}
\label{shape}
Y(K) = Y_{\infty}\frac{(K/\alpha)^{3/2}-1}{(K/\alpha)^{3/2}+\beta}\,.
\end{equation}
For $\beta\gg K/\alpha$,  Eq.~\ref{shape} describes the field ionization behavior, $Y(K) \propto K^{3/2}$,
while for $\beta \ll K/\alpha$,  the typical DII shape emerges with a sharp onset followed by saturation at $Y_{\infty}$. The latter is characterized by a negative second derivative
$Y''(K)$ while for the field ionization power law $Y''(K)>0$ holds for all $K$.
Hence, $Y''(K_\mathrm{cr})=0$  can be interpreted as the conditional equation for the critical number of dopant atoms where
field ionization dominated absorption goes over into DII dominated absorption.  Using Eq.~\ref{shape},
we get $K_\mathrm{cr} = \alpha(\beta/5)^{2/3}$. The parameters $(\alpha,\beta)$ are obtained
for each pulse length $T$ by fitting Eq.~\ref{shape} to the numerically generated data
in Fig.~\ref{fig:PulseLength}.  $K_\mathrm{cr}$ exhibits an exponential dependence on the pulse length $T$, as can be seen from Fig.~\ref{fig:field2ignition}.  The shaded area indicates DII with $Y'' < 0$, while below the line of $K_\mathrm{cr}$ field ionization rules. Since physically, at least one dopant atom is necessary for ionization, DII dominates for $T$ larger than about 9\,fs. This is consistent with the qualitative conclusion from Fig.~\ref{fig:PulseLength} as well as with our experimental result which demonstrates DII for a pulse length of 10\,fs.

In summary, we have experimentally demonstrated that it is possible to transfer energy resonantly from a 790 nm intense few-cycle pulse to bound electrons without the need of atomic expansion of the target. This brings down the time scale of this extremely efficient but relatively slow process from the sub-picosecond regime down to a few femtoseconds and, at the same time, allows it to be applied to any form of matter that can be suitably doped with seed atoms. Finally, the dopant induced ignition demonstrated here, may also explain the surprising enhancement of light absorption in water doped argon clusters \cite{jhmk08}.

Stimulating discussions with J. Tiggesb{\"a}umker and Th. Fennel, and the valuable assistance of O. B{\"u}nermann in the Monte-Carlo calculations are gratefully acknowledged. This work is supported by the Deutsche Forschungsgemeinschaft.


%

\end{document}